# Plasmon-emitter interaction using integrated ring grating-nanoantenna structures

**Nancy Rahbany[1], Wei Geng[1], Renaud Bachelot[1] and Christophe Couteau[1,2,3]**

[1] Laboratory of Nanotechnology, Instrumentation and Optics, ICD CNRS UMR 6281, University of Technology of Troyes, 10000, Troyes, France
[2] CINTRA CNRS-Thales-NTU, UMI 3288, Research Techno Plaza, 50 Nanyang Drive, Singapore
[3] Centre for Disruptive Photonics Technologies (CDPT), Nanyang Technological University, Singapore

E-mail: christophe.couteau@utt.fr

**Abstract.** Overcoming the diffraction limit to achieve high optical resolution is one of the main challenges in the fields of plasmonics, nanooptics and nanophotonics. In this work, we introduce novel plasmonic structures consisting of nanoantennas (nanoprisms, single bowtie nanoantennas and double bowtie nanoantennas) integrated in the center of ring diffraction gratings. Propagating surface plasmon polaritons (SPPs) are generated by the ring grating and coupled with localized surface plasmons (LSPs) at the nanoantennas exciting emitters placed in their gap. SPPs are widely used for optical waveguiding but provide low resolution due to their weak spatial confinement. Oppositely, LSPs provide excellent sub-wavelength confinement but induce large losses. The phenomenon of SPP-LSP coupling witnessed in our structures allows achieving more precise focusing at the nanoscale, causing an increase in the fluorescence emission of the emitters. FDTD simulations as well as experimental fabrication and optical characterization results are presented to study plasmon-emitter coupling between an ensemble of dye molecules and our integrated plasmonic structures. A comparison is given to highlight the importance of each structure on the photoluminescence and radiative decay enhancement of the molecules.

## 1. Introduction

Researchers are constantly exploring new ways to improve the fabrication of micro and nano-optical devices capable of controlling and enhancing surface plasmon launching, propagation, and localization(1). However, the challenge in these devices resides in the confinement of light into sub-wavelength regions which is limited by diffraction. It has been shown that both surface plasmon polaritons (SPPs) and localized surface plasmons (LSPs) are indispensable components for optical applications at the nanoscale. SPP waveguiding and confinement can be achieved by several ways including the simple configuration of a thin metal film sandwiched between two symmetric dielectric layers(2), gaps and V-grooves(3,4), near field optical sources(5,6), stripes and nanowires(7–9), near-field coupling between adjacent metallic nanoparticles in linear chains(10), and metallic gratings(11–13). Optical nanoantennas, on the other hand, benefit from their sizes, which are comparable or smaller than the wavelength of visible light, to overcome the diffraction limit and manipulate electromagnetic fields at the nanoscale(14,15). As a result, they are widely used in many applications such as near-field optical microscopy(16), surface enhanced spectroscopy(17), sensing(18), medical therapy(19) and optoelectronic devices(20).

Combining diffraction gratings with nanoscale apertures and nanoantennas benefits from the efficient coupling between SPPs and LSPs to create highly confined, enhanced, and collimated electromagnetic fields(21–29). However, a thorough study on the influence of such integrated structures on the fluorescence enhancement of emitters still lacks. In our previous work(30), we analyzed experimentally and numerically the directional launching and detection of SPPs using a plasmonic platform consisting of a gold ring grating. SPP-emitter coupling was studied by exciting fluorescent molecules placed in the center of the rings. In another work(31), we presented numerical characterizations of the effect of the double bowtie geometry on the electromagnetic field enhancement in its gap due to localized surface plasmons. In this work, we present the combination of both structures via two plasmonic devices responsible for focusing and enhancing electromagnetic fields at the nanoscale even further. The first consists of nanoantennas integrated in the center of ring diffraction gratings, where SPPs are generated by the ring grating and couple to LSPs at the nanoantennas. The second structure consists of a double cavity containing a ring grating and a nanoantenna. For both structures, the enhanced electromagnetic field in the nanoantenna gap leads to the excitation of dye molecules causing an increase in their fluorescence and a decrease in lifetime. FDTD simulations and photoluminescence spectra are performed on rings containing nanoprisms, single bowtie nanoantennas and double bowtie nanoantennas, illustrating that for both types of structures (structure 1 with bigger rings and structure 2 with smaller rings to form a double nanocavity), double bowtie nanoantennas lead to the highest fluorescence enhancement. Measurements on rings of different diameters allow us to extract the propagation length. Finally, a comparison between the two structures is given, showing that the radiative decay rate enhancement is approximately the same for antennas with big gap sizes (*100 nm*). However, for smaller gaps (*50 nm*), the double cavity structure starts having a more significant effect on the fluorescence of the emitters with a radiative decay rate enhancement of 6.8. Even though even smaller gaps could be more beneficial, we think that this is a good compromise considering the difficulty of fabricating smaller structures and for future coupling and manipulating with single emitters.

## 2. Structure Description

### 2.1. Nanoantenna in the center of ring grating (structure 1)

The first structure is composed of a gold ring grating made of 5 concentric circular grooves with a nanoantenna placed in its center, as seen in the inset of figure 1. The nanoantennas we choose to study are a nanoprism, a single bowtie and a double bowtie. Upon illuminating the ring grating with a laser source on the circumference, SPPs are generated and propagate to get focused in the center(32). When the laser source is placed at a position facing a triangle side, SPPs get directed along the two other sides of the triangle and form an electromagnetic hotspot in the nanoantenna gap. We first start by performing some FDTD numerical simulations, using Lumerical software, to measure the electric field intensity in the gap of a gold nanoantenna placed in the center of a ring grating, all on a gold substrate.

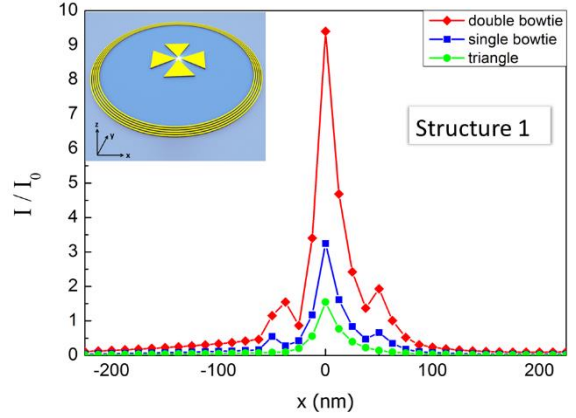

Figure 1. Structure 1: Electric field intensity enhancement as a function of the position along the x-axis (with y being at the position of the nanoprism tip) for rings with a nanoprism, a single bowtie and a double bowtie nanoantenna placed in the center. Inset: Schematic of a double bowtie nanoantenna placed in the center of a ring grating structure.

The dimensions of the structure are chosen to satisfy the conservation of momentum relation:

$$\frac{2\pi}{\lambda}n_{eff} = \frac{2\pi}{\lambda}\sin\theta + \frac{2\pi}{d} \qquad (1)$$

where $\lambda$ is the wavelength of the incident excitation light source, $n_{eff}$ is the effective refractive index, $\theta$ is the angle of incidence, and $d$ is the grating period.
In our case, we take the incident angle to be $\theta = -10°$ (optimized angle of incidence for such grating structures calculated in our previous article(30)), and the wavelength of the laser used is $\lambda = 632.8$ *nm* that is compatible for exciting Atto-633 dye molecules placed in the nanoantenna gap (which will be presented in the experimental results in the next section). The SPP effective index on an air-gold interface at $\lambda = 632.8$ *nm* is $n_{eff} = 1.0459 + 0.0069\,i$, and therefore the period is calculated to be $d = 519$ *nm*. A linear plane wave polarized perpendicular to the grating grooves (TM polarization) is incident at a certain position on the ring grating circumference. A nanoantenna is positioned exactly in the center of a *10 μm* ring with the triangle side facing the grating grooves where the incident excitation is placed. The triangle side length is chosen to be *2 μm* and its thickness *120 nm*. Perfectly matched layer (PML) boundaries are used to absorb incident electromagnetic waves and avoid reflections. A "frequency-domain field and power" monitor is placed on the surface of the grating and nanoantenna, i.e. at a height of *120 nm*, to record the electric field intensity along the *x*-direction and for *y* being at the position of the nanoantenna gap. Due to the random distribution of molecules in the gap, and since the value of the electromagnetic field linearly increases with increasing height, the position of the monitor is chosen at the top surface to record the maximum excitation intensity which predominantly contributes to the PL enhancement measured experimentally. The recorded values are normalized by the intensity of the incident light source, resulting in the electric field enhancement created in the gap of each structure ($I/I_0$). We compare the values of the intensities at the tip of the nanoprism to that in the *100 nm* gaps of single and double bowtie nanoantennas. The results are displayed in figure 1 where we observe that the double bowtie nanoantenna leads to the highest electromagnetic confinement in the gap. The *x*-axis in this figure refers to the position along the x-direction in the simulations (see inset of figure 1). Since the gap size is quite big, two additional peaks appear at about *50 nm* away from the position of the center of the gap, corresponding to intensity hotspots created at the triangle tips.

*2.2. Double cavity (structure 2)*
We then perform the same type of simulations but for the cavity structure presented in figure 2.

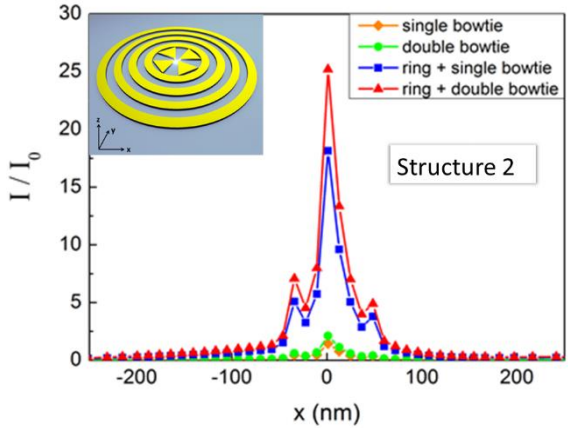

**Figure 2.** Structure 2: Electric field intensity enhancement as a function of the position along the x-axis (with y being at the position of the nanoantenna gap) for single and double bowtie nanoantennas, and cavities with single and double bowties. Inset: Schematic of a double cavity with a double bowtie nanoantenna.

In this structure, both the grating and the nanoantenna act as one plasmonic cavity that concentrates the incident field and excites emitters placed in the gap. A circularly polarized light source (addressing both the horizontal and vertical components of the structure) is now made incident on a double cavity composed of a ring grating of 5 concentric grooves separated by a period *d = 519 nm* containing a single or double bowtie nanoantenna of *2 μm* side length. The electric field intensity is recorded in the gap of the integrated cavities and compared to that obtained from the nanoantennas alone. The highest intensity is observed for the double cavity with a double bowtie, as seen in figure 2, with *x* also being the position along the *x*-direction in the simulations.

## 3. Fabrication and Optical Characterization

*3.1. Nanoantenna in the center of ring grating (structure 1)*

To test our structures experimentally, we fabricated using electron beam lithography (EBL) (e-beam dose = *90 μC/cm$^2$*), ring gratings with a period of *519 nm* containing nanoantennas (nanoprisms, single bowtie nanoantennas, and double bowtie nanoantennas) in their centers. Rings with *5 μm*, *10 μm*, *20 μm* and *30 μm* diameters, and nanoantennas with *1 μm* and *2 μm* side lengths and *50 nm*, *100 nm* and *150 nm* gaps were fabricated on silicon substrates. After the EBL process, a *120 nm* layer of gold is evaporated and kept on the structures as well as inside the rings to allow SPP propagation. The optical and SEM images in figure 3a and figure 3b respectively show ring gratings of *10 μm* diameters containing nanoantennas of *2 μm* side lengths and *100 nm* gaps. A homogeneous layer of Atto-633 dye molecules (concentration = *3.33 mg/L*) is spin coated on the structures. Those molecules act as probes for plasmonic imaging of SPPs as well as candidates for studying plasmon-emitter coupling and enhancing their emission properties. They are observed under a home-built confocal microscope system of high sensitivity including a *50X*, *NA=0.95* microscope objective, a spectrometer, with a Peltier cooled CCD camera at *T = -80$^o$C*. In order to test the homogeneity, the photoluminescence (PL) spectrum is measured at several places on the surface by exciting the molecules with a *632.8 nm* continuous diode laser, which resulted in an identical spectrum for all locations (with less than 5% change).

To study SPP propagation in our structure, we excite ring gratings with nanoprisms of 1 μm side lengths in the center by placing the laser spot on the ring circumference at a position facing the nanoprism side (see figure 4a). We observed that this is the optimized configuration for the laser spot where SPPs are generated at the ring and propagate towards the nanoprism and form an intense electromagnetic hotspot at the tip. This can be seen in figure 4b that displays the PL spectra of the dye molecules at different locations on the prism, where the highest corresponds to the dyes on the tip. A *10 nm* cut is observed in the experimental curve around *633 nm* due to a notch filter placed at the output to eliminate any light coming from the laser.

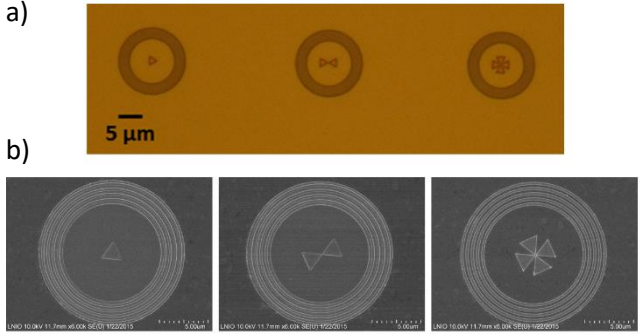

**Figure 3.** a) Optical and b) SEM images of gold nanoantennas in the center of ring gratings on a Si substrate.

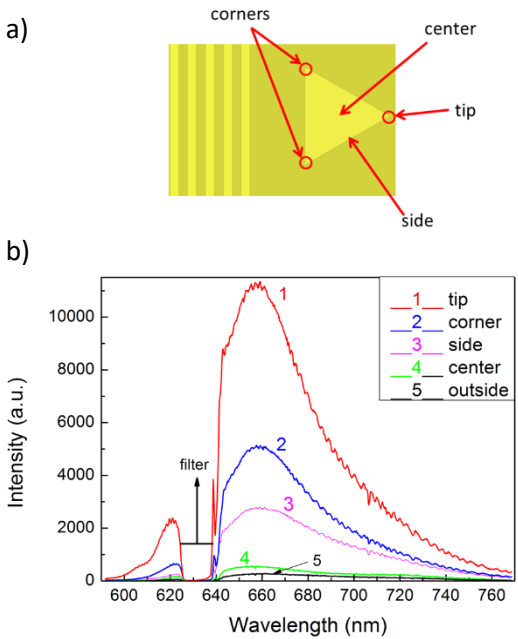

**Figure 4.** Electromagnetic confinement at the nanoprism tip. a) Schematic of the different measurement locations, b) PL spectra of Atto-633 dyes at the tip, corner, side and center of the nanoprism.

To measure the SPP propagation length in our structures, we record the PL spectra for dye molecules placed at the nanoprism tips in rings of different diameters ranging from 5 *μm* - 30 *μm*. The results are shown in figure 5 where the PL intensity is plotted for four different diameters (figure 5a) and as a function of the distance travelled by SPPs (figure 5b). The data in figure 5b is recorded at the emission wavelength of the dye molecules; $\lambda_{em} = 657$ *nm*. The fit of the exponential curve results in the measured experimental value of the propagation length $L_{exp} = 19.9 \pm 0.05$ *μm*. The expected value obtained from FDTD numerical simulations is $L_{sim} = 17.9$ *μm*. These two results are compatible however they slightly surpass the values obtained in the literature(33) ($L_{lit} = 10$ *μm*), which indicates that our configuration is successful in launching surface plasmons to a further distance away from the grating.

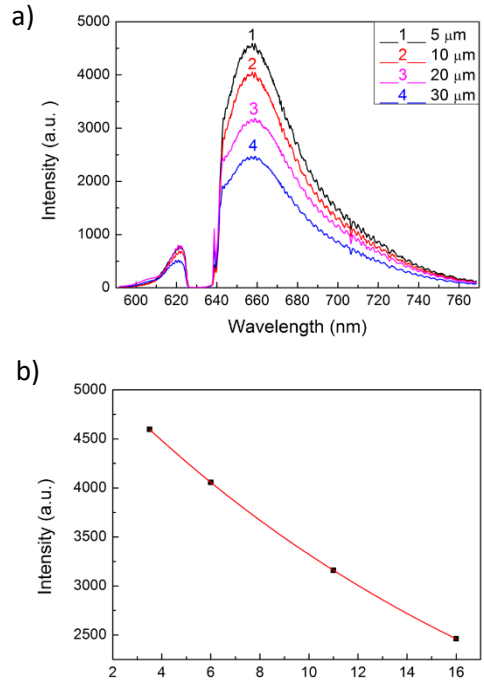

**Figure 5.** Propagation length of SPPs on a gold-air interface at $\lambda = 633$ *nm*. a) PL spectra of Atto-633 dyes at the nanoprism tip for ring gratings of 5 μm, 10 μm, 20 μm, and 30 μm diameters, b) PL spectra of Atto-633 dyes at the nanoprism tip as a function of the distance travelled, measured at $\lambda$em = 657 nm.

We then perform PL and lifetime measurements on dye molecules placed in the gap of single and double bowtie nanoantennas in the center of ring gratings of *10 μm* diameters using a *640 nm* pulsed laser of *3.07 mW* power. The laser spot position is maintained on the ring circumference facing the nanoantenna side. The hotspot created by the propagating SPPs excites the dye molecules placed in the nanoantenna gap for which we measure the PL intensity and lifetime. The lifetime is measured using a time correlated single photon counting setup (TCSPC) connected to our home-made confocal microscope. A comparison between dyes placed on the gold substrate outside the structures, on a nanoprism tip, and in the gap of single and double bowtie nanoantennas is given in figure 6, where we can see that the highest PL intensity ($P_l$) (figure 6a) and lowest lifetime ($\tau$) (figure 6b) correspond to dyes in the gap of double bowtie nanoantennas ($\tau_0/\tau = 1.54$ and $P_l/P_{l,0} = 13.7$).

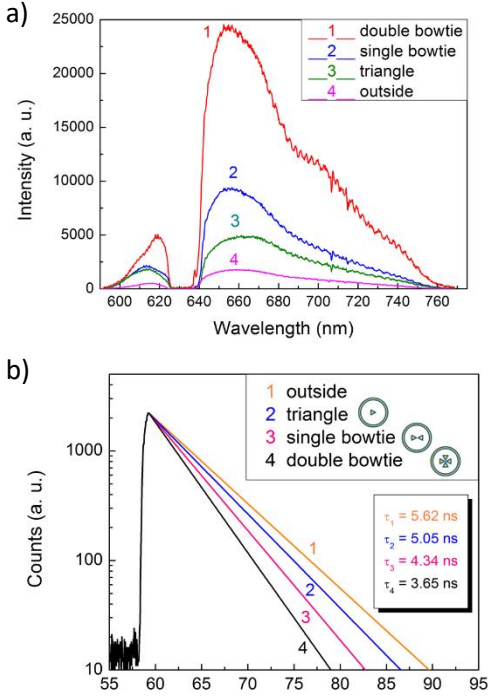

**Figure 6.** Photoluminescence and lifetime of Atto-633 dyes on nanoantennas in the center of ring gratings. a) Highest obtained PL spectra as a function of wavelength and b) lifetime of dyes measured outside the structures, at the tip of a nanoprism, in the gap of a single bowtie nanoantenna and in the gap of a double bowtie nanoantenna.

### 3.2. Double cavity (structure 2)

We then study the second type of structures composed of a double cavity. For that, we fabricated single and double bowtie nanoantennas of *1 µm* and *2 µm* side lengths surrounded by ring gratings of *d = 519 nm* periods. The same EBL process is followed and a homogeneous layer of Atto-633 is also spread on the surface. In figure 7a and figure 7b respectively, we show optical and SEM images of *2 µm* sided single and double bowtie nanoantennas with a *100 nm* gap, as well as double cavities containing these structures. For these structures, we excite the dye molecules by placing the laser spot centered on the nanoantenna gap (*785 µW* power). PL and lifetime spectra are then simultaneously measured. A comparison is done between the emission of dyes outside the structures, in the gap of single and double bowtie nanoantennas, and in the cavities with single and double bowtie nanoantennas (figure 8). As expected, cavities containing double bowtie nanoantennas lead to the highest PL intensity and lowest lifetime ($\tau_0/\tau = 1.47$ and $P_l/P_{l,0} = 31.3$).

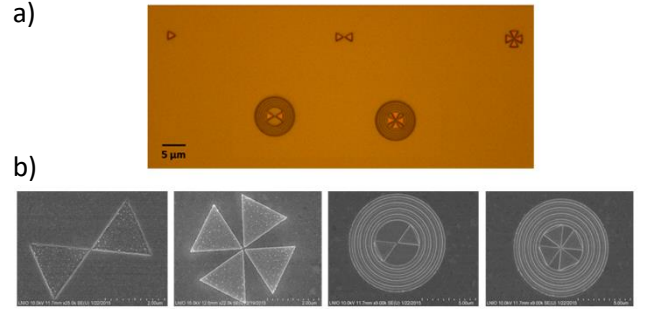

**Figure 7.** a) Optical and b) SEM images of gold nanoantennas and double cavities engraved on a Si substrate.

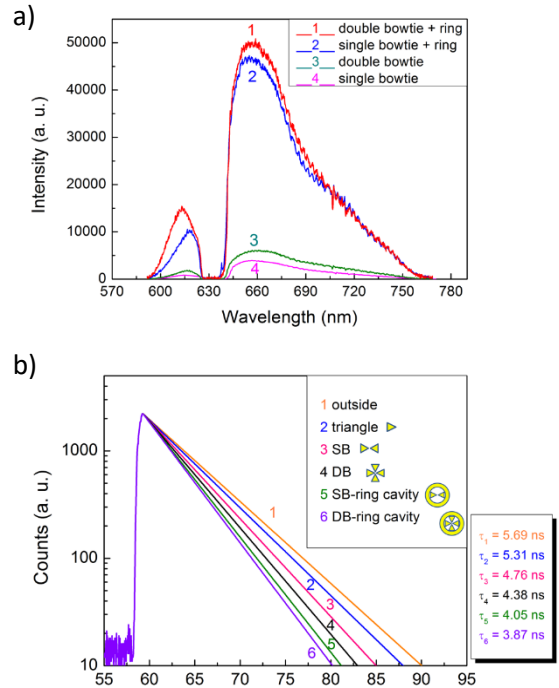

**Figure 8.** Photoluminescence and lifetime of Atto-633 dyes in the gap of double cavities. a) Highest obtained PL spectra as a function of wavelength and b) lifetime of dyes measured outside the structures, at the tip of a nanoprism, in the gap of a single bowtie nanoantenna (SB), double bowtie nanoantenna (DB), cavity with single bowtie, and a cavity with double bowtie. (see figure 7)

### 4. Plasmon-Emitter Coupling Analysis

Experimentally, the fluorescence enhancement, i.e. the ratio between the measured fluorescence intensity of the emitters in the nanoantenna and its intrinsic value outside the nanoantenna ($P_l/P_{l0}$) can be expressed in terms of the ratios of the excitation rate ($R_{exc}/R_{exc,0}$), the emitter's quantum yield ($\eta/\eta_0$), and on the collection efficiency of the setup ($C_{coll}/C_{coll,0}$)(34):

$$\frac{P_l}{P_{l,0}} = \frac{R_{exc}}{R_{exc,0}} \cdot \frac{\eta}{\eta_0} \cdot \frac{C_{coll}}{C_{coll,0}} \quad (2)$$

The excitation rate can be expressed as:

$$R_{exc} = I_{exc} |\mu_{12}|^2 \langle \cos(\alpha) \rangle^2 \quad (3)$$

where $I_{exc}$ is the local excitation intensity, and $\vec{\mu}_{12}$ and $\alpha$ are respectively the molecule's electric dipole moment and orientation.
Taking the quantum yield as $\eta = \tau \cdot \Gamma_r$ with $\tau$ and $\Gamma_r$ the lifetime and radiative decay rate, and averaging over the random orientation of the molecules; $<\cos(\alpha)>^2 = 1/2$, Eq. 2 becomes:

$$\frac{P_l}{P_{l,0}} = \frac{I_{exc}}{I_{exc,0}} \cdot \frac{\tau}{\tau_0} \cdot \frac{\Gamma_r}{\Gamma_{r,0}} \cdot \frac{C_{coll}}{C_{coll,0}} \quad (4)$$

To compare the effect of structure 1 and structure 2 on the radiative decay rate enhancement of the dye molecules ($\Gamma_r/\Gamma_{r,0}$), we make use of the values of the lifetime reduction ($\tau_0/\tau$) and PL intensities ($P_l/P_{l,0}$) obtained experimentally from both structures, as well as the excitation intensity from the FDTD simulations ($I_{exc}/I_{exc,0}$). The results are given in Table 1. We take the collection efficiency term ($C_{coll}/C_{coll,0}$) to be on the order of 1 due to the high numerical aperture of the objective used ($NA = 0.95$), which is shown to collect 98.5% of the emitted light. This was obtained by numerical calculations taking into account the emission angle for dipoles of different orientations emitting at $\lambda = 657~nm$ on a gold substrate. Therefore, after plugging in those values into Eq. 4, we realize that both structures 1 and 2 (with *100 nm* gaps) lead to approximately the same quantum efficiency and radiative decay rate enhancements. This result is not intuitive especially after observing a much higher PL enhancement in structure 2 while the change in lifetime is nearly the same. However, it can be explained by the fact that the PL enhancement observed experimentally is due to the increase in the local excitation intensity caused by the structures, which is higher for structure 2. Several reasons might explain the similar reduction of lifetime; the nanoantenna gap is quite big (*100 nm*), Atto-633 dyes have a high intrinsic quantum efficiency (64%) which gives a lower chance for observing high radiative decay rate enhancement, emitters might be deviated from the maximum field in the gap, and the dipole moments of the emitters might not be fully aligned with the field.

In order to start observing a difference between the two structures, we must study antennas with smaller gap sizes. We carried out some numerical simulations and PL measurements on similar structures with *50 nm* gaps, which showed a higher radiative decay rate enhancement for emitters placed in structure 2 as compared to structure 1. The results are presented in figure 9 where the simulated electric field intensity (figures 9a,b) and the experimentally measured PL intensity of dye molecules (figures 9c,d) are measured and compared to structures with *100 nm* gaps. The values are summarized in Table 1 where we notice that structure 2 now causes a bigger decrease in lifetime as opposed to structure 1 (also shown in figure 9e). This leads to a more significant increase in the radiative decay rate enhancement ($\Gamma_r/\Gamma_{r,0} = 6.8$ for structure 2 and $\Gamma_r/\Gamma_{r,0} = 2.2$ for structure 1), i.e. a higher Purcell enhancement. We also notice that the gap size has no significant impact on the radiative decay rate enhancement of structure 1. Therefore, we conclude that as we go towards smaller gaps, structure 2 appears to have a stronger influence on the fluorescence enhancement of emitters placed in its gap, due to a stronger SPP-LSP interaction. This is caused by the double cavity effect, where a much higher electric field is formed in the gap which in turn significantly enhances the radiative emission of the emitters and reduces their lifetime. While on the other hand, structure 1 appears to be more beneficial in guiding SPPs and can therefore be more efficient in coupling to waveguides or addressing specific nanostructures.

**Table 1.** Recorded values of the ratios of the PL intensity ($P_l/P_{l,0}$), lifetime ($\tau_0/\tau$), excitation intensity ($I_{exc}/I_{exc,0}$), quantum efficiency ($\eta/\eta_0$), and radiative decay rate ($\Gamma_r/\Gamma_{r,0}$) obtained for structures 1 and 2 with *100 nm* and *50 nm* gap sizes.

|  | 100 nm gap | | 50 nm gap | |
| --- | --- | --- | --- | --- |
|  | Str. 1 | Str. 2 | Str. 1 | Str. 2 |
| $P_l/P_{l,0}$ | 13.7 | 31.3 | 14 | 49 |
| $\tau_0/\tau$ | 1.54 | 1.47 | 2.15 | 4.75 |
| $I_{exc}/I_{exc,0}$ | 9.4 | 25.2 | 13.9 | 34 |
| $\eta/\eta_0$ | 1.4 | 1.2 | 1.01 | 1.4 |
| $\Gamma_r/\Gamma_{r,0}$ | 2.2 | 1.8 | 2.2 | 6.8 |

## 5. Conclusions

In this work, we proposed two types of integrated ring grating/nanoantenna structures that are used to improve the localization and intensity of electromagnetic fields at the nanoscale. Experimental observations and FDTD numerical simulations demonstrated that structures containing double bowtie nanoantennas lead to the highest field confinement as opposed to the other nanoantennas studied (single bowties and nanoprisms). The propagation length was calculated which surpasses what is obtained in the literature so far. We also showed how these structures lead to the enhancement of the photoluminescence and lifetime of emitters placed in their gaps. Theoretical calculations were given to discuss plasmon-emitter coupling in the weak coupling regime, and compare the effect of both structures on the fluorescence enhancement of the dye molecules. We observed that for big gaps (*100 nm*), both structures lead to approximately the same radiative decay rate enhancement, despite the higher PL intensity created in structure 2. This is due to the fact that the PL enhancement is solely caused by the excitation rate enhancement in the gap. However for small gaps (*50 nm*), structure 2 starts showing a higher radiative decay rate enhancement and a bigger lifetime reduction. Therefore, this work reinforces the fact that an increase in the photoluminescence alone is not sufficient for characterizing plasmonic nanoantennas(34). A thorough study of the coupling between SPPs and LSPs in our structures was presented, allowing us to conclude that structure 1 is beneficial in directing SPPs on metallic surfaces which can be used in numerous applications such as addressing specific structures on the surface or coupling to waveguides. Structure 2, on the other hand, is shown to have a more important effect on the Purcell enhancement of emitters due to a more efficient SPP-LSP coupling. Therefore, depending on the desired outcome, the design of the structures can be carefully chosen to include propagating surface plasmons, localized surface plasmons, or the coupling of both, providing flexibility in addressing particular applications.

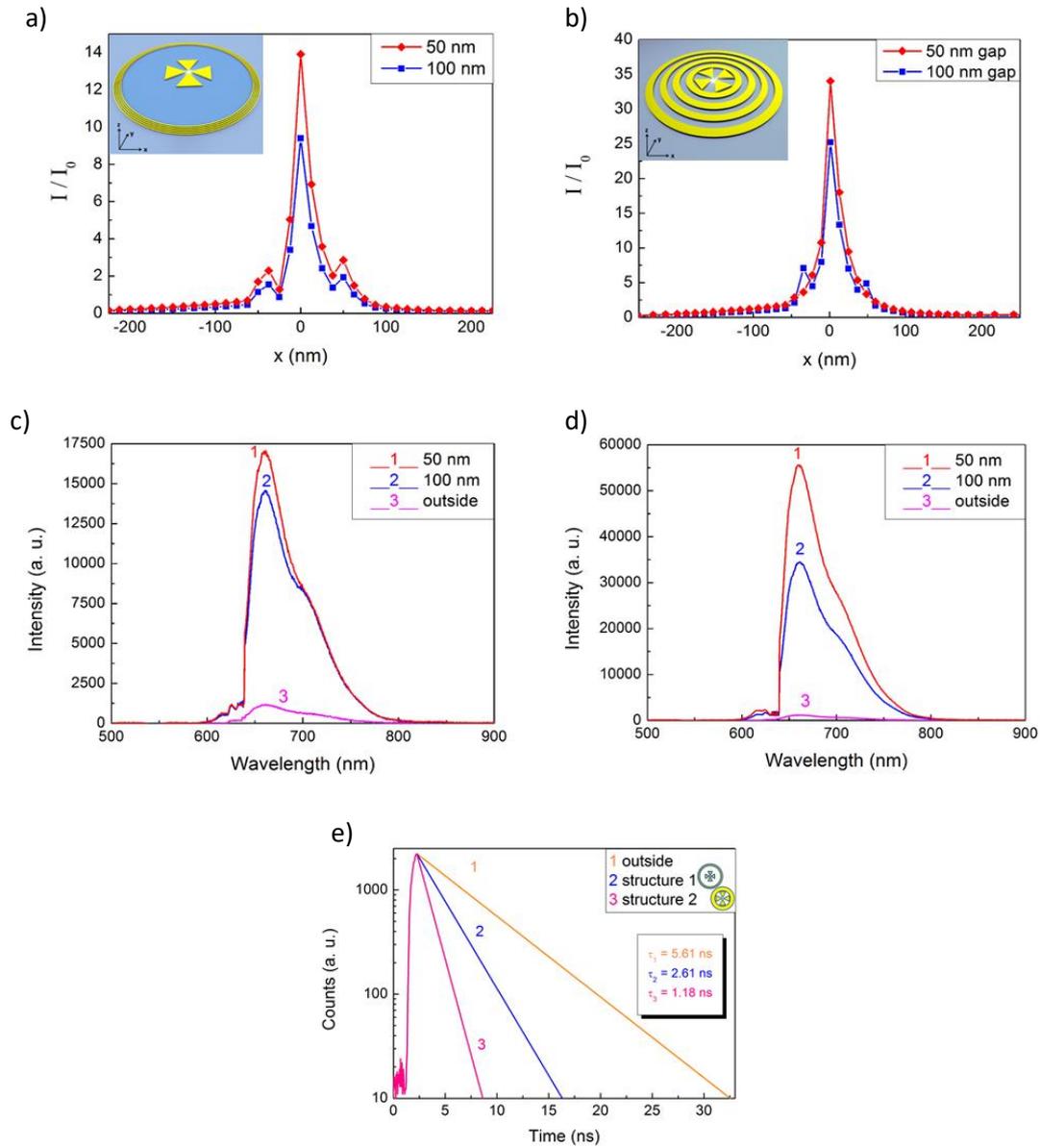

**Figure 9.** Comparison between structures having *50 nm* and *100 nm* gaps: a-b) FDTD simulations of the electric field intensity enhancement in the gap of double bowtie nanoantennas of *50 nm* and *100 nm* gaps for a) structure 1 and b) structure 2, c-d) PL measurements of Atto-633 dyes placed in the gaps of the structures presented in a) and b), e) Lifetime measurements of Atto-633 dyes placed in the gap of structures 1 and 2 with *50 nm* gaps.

Further work can be done with our integrated plasmonic structures. This includes studying the effect of having even smaller gap sizes aiming to obtain a higher Purcell factor and a more significant change in lifetime. In addition, a far-field emission characterization study can be done which enables the control over the direction of emission of the emitters. Another attempt would be to increase the concentration of dye molecules in the gap which facilitates reaching the strong coupling regime(35,36). An additional study includes coupling single photon sources to our plasmonic structures which is expected to increase their collection and emission efficiencies. Therefore, we show that the efficient coupling between propagating surface plasmons and localized surface plasmons present in our structures allows us to achieve high electromagnetic confinement at the nanoscale, which can be used to increase the fluorescence emission of an ensemble of emitters as well as single emitters placed in their vicinity.

**Acknowledgements**

N. R. would like to thank the French Ministry of Education for her PhD grant. R. B. and C. C. would like to acknowledge the financial support of the Labex Action program and the COST program "Nanoscale Quantum Optics-NQO". The authors thank the region Champagne-Ardenne platform Nanomat for fabrication and characterization facilities.